M. Eng. Wojciech Sadowski[1,*],
M. Sc. Christin Velten[2],
M. Sc. Maximilian Brömmer[3],
M. Sc. Hakan Demir[1],
M. Sc. Kerstin Hülz[2],
Prof. Dr. Francesca di Mare[1],
Dr.-Ing. habil. Katharina Zähringer[2],
Prof. Dr.-Ing. Viktor Scherer[3]

# Low Reynolds number flow in a packed bed of rotated bars

The present study focuses on the gas flow through an experiment-scale modular packed bed reactor consisting of square bars, arranged in layers. Each layer is rotated by 30°, resulting in a complex shape of the void spaces. Particle Image Velocimetry measurement results inside and above the bed are presented for Reynolds numbers of 100 and 200, and used as validation data for two sets of particle-resolved numerical simulations, using a boundary-conforming meshing strategy and treating the solid boundaries via the blocked-off method. The flow inside the bed is largely independent of the Reynolds number and determined by the void space geometry. The flow in the freeboard is dominated by the presence of slowly dissipating jets downstream of the bed. The numerical results are in good agreement with the measurements, both inside and above the bed; however, stronger deviations can be observed in the freeboard and can be traced to numerical properties of the current simulation approaches.

**Author affiliations**

[1]Chair of Thermal Turbomachinery and Aero Engines, Ruhr University Bochum, Universitätsstraße 150, D-44801, Bochum, Germany.

[2]The Laboratory of Fluid Dynamics and Technical Flows, Otto-von-Guericke-Universität, Universitätsplatz 2, 39106, Magdeburg, Germany.

[3]Institute of Energy Plant Technology, Ruhr University Bochum, Universitätsstraße 150, D-44801, Bochum, Germany.

Email corresponding author: wojciech.sadowski@rub.de

ORCID iDs of the authors: Sadowski: 0000-0002-1090-9109, Brömmer: 0000-0002-8650-5776, Demir: 0009-0006-3929-8352, Velten: 0009-0008-1243-729X, Zähringer: 0000-0002-4780-6023

## 1 Introduction

In packed-bed reactors, an assembly of particles is typically passed by a gas introducing reactants or facilitating processes, e.g., drying or coating. Unfortunately, *a priori* estimation of the flow conditions in between the particles is a difficult task, complicating reactor design and leading to large error margins. Although the packing often consists of randomly shaped polyhedra or other complex particle shapes [1], a typical simplification in both analytical and numerical models of packed beds is the assumption of spherical particles [2, 3], contrary to observations that the particle shape influences the flow field to a large extent [4].

The most accurate way to predict the movement of the gas phase *in silico* involves the use of Computational Fluid Dynamics (CFD) simulations, in particular the particle-resolved approach [5], which resolves both, the geometry of the packing and the flow topology. The method offers high fidelity at a large com-putational cost, limiting its use to laboratory-scale configurations [5, 6]. Moreover, when a standard Finite Volume Method (FVM) is used for such simulation, often a need to locally modify the geometry arises to treat the regions of poor numerical stability near the contact regions between individual particles [5].

Immersed boundary method (IBM) [7] addresses these challenges by avoiding boundary-conforming mesh generation, implicitly treating contact points [8], and eliminating the need for expensive remeshing with moving objects. One such approach is the blocked-off method, proposed by Patankar [9], which does not require complex interpolation or surface reconstruction. This method has been validated against experimental data and smooth IBM for packed bed simulations [10], has been applied to study pressure drop in beds of spherical particles [11] and to model locally resolved flow near reactor fuel lances [12]. Its advantages include the direct application of wall functions at higher Reynolds numbers and straightforward integration into flow solvers.

Importantly, CFD predictions require validation against flow measurements, but such data are rarely available for packed bed reactors—often only macroscopic quantities are being observed, e.g. pressure drop [13]. To measure the flow inside the bed, intrusive measurement techniques are often of limited use as they can disrupt the flow field. Although, enabling optical access into the bed is often difficult in practice, it can be achieved *via* the use of transparent packing materials and refractive index matching (RIM), often along with additional post-processing. RIM-based approaches have been successfully used for, among others, Particle Tracking Velocimetry [14], planar Particle Image Velocimetry (PIV) [15, 16] and tomographic PIV [17]. Non-intrusive methods not requiring optical access (e.g. MRI), are typically only limited to liquids and lack the resolution of optical approaches [18].

Recently, Velten and Zähringer [19] used PIV to measure both the flow in the interstices of a model packed bed with a BCC packing of spheres and the conditions above the bed. Although, the interstitial data was gathered in a set of small, optically accessible regions, it enabled a detailed analysis of the flow field and validation of several methods and numerical codes [6,10]. These applications highlight the synergies of combining experimental and numerical analyses for the study of flow conditions in packed beds.

Moving away from the typically assumed spherical particle geometry, in the present work, we investigate the flow in a packed bed formed from the assembly of square bars. The configuration results in interstices with complex geometries with multiple inlets and outlets. The flow field is investigated experimentally, using PIV, and numerically, by two different particle-resolved FVM-based simulation methods: the incompressible solver available in OpenFOAM-12 package, employing a boundary-conforming mesh, and the blocked-off approach implemented in the in-house Discrete Element Method (DEM) extension to OpenFOAM-v2012. Two different Reynolds numbers are studied, and for each case, results are compared both within the bed and in the freeboard region.

## 2 Packed-bed geometry and flow conditions

The packed bed geometry (see Fig. 1a) is based on a modular design, where the bed is assembled by stacking the modules on each other as shown in Fig. 1(b). Each module consists of five parallel bars (square cross-section with size $B = 10$ mm) spaced with a distance of $5$ mm inside dodecagon-shaped side walls. The outer walls of the modules are circular, allowing for each module to be rotated freely. This enables the study of various regular or arbitrary geometrical configurations at a constant porosity of $\phi = 0.332$, defined as a fraction of the void space to the total module volume $BA$, where $A$ is the cross-section area of $3000\ mm^2$. In the current study, the bed consists of a total of 18 modules assembled with the rotation angle of $30°$ as indicated in Fig. 1.

The flow is characterized by a particle Reynolds number $\mathrm{Re}_p = \langle w \rangle B/\nu$, based on the intrinsic average of the velocity of the gas flowing through the packing $\langle w \rangle$ and the bar size $B$. The chosen Reynolds numbers of $\mathrm{Re}_p = 100$ and $200$ correspond to inertial flow conditions, which span both steady-state and transient flow. The velocity $\langle w \rangle$ can be computed from the cross-section area $A$ and the volumetric flow rate of the device $Q = A\phi\langle w\rangle$. For Reynolds numbers of 100 and 200 these velocities are 0.151 and 0.303 m/s, respectively.

Moreover, the choice of low Reynolds numbers enables detailed comparison of the performance and accuracy of the numerical codes in steady and unsteady flows, as the resolution requirements for Direct Numerical Simulation (DNS) can be reached without exceedingly high computational cost. At $\mathrm{Re}_p = 200$, although the flow can be expected to be unsteady, with strong inertial effects (as indicated by the work of Demir et al. [20] focused on the pressure drop in the same packed bed geometry) it remains predominantly laminar with possible intermittent fluctuations. Based on established literature [21], well defined turbulence can be established above $\mathrm{Re}_p \sim 750$ (interested reader is also referred to [22] where measurements for $\mathrm{Re}_p = 1000$ in the studied configuration are available).

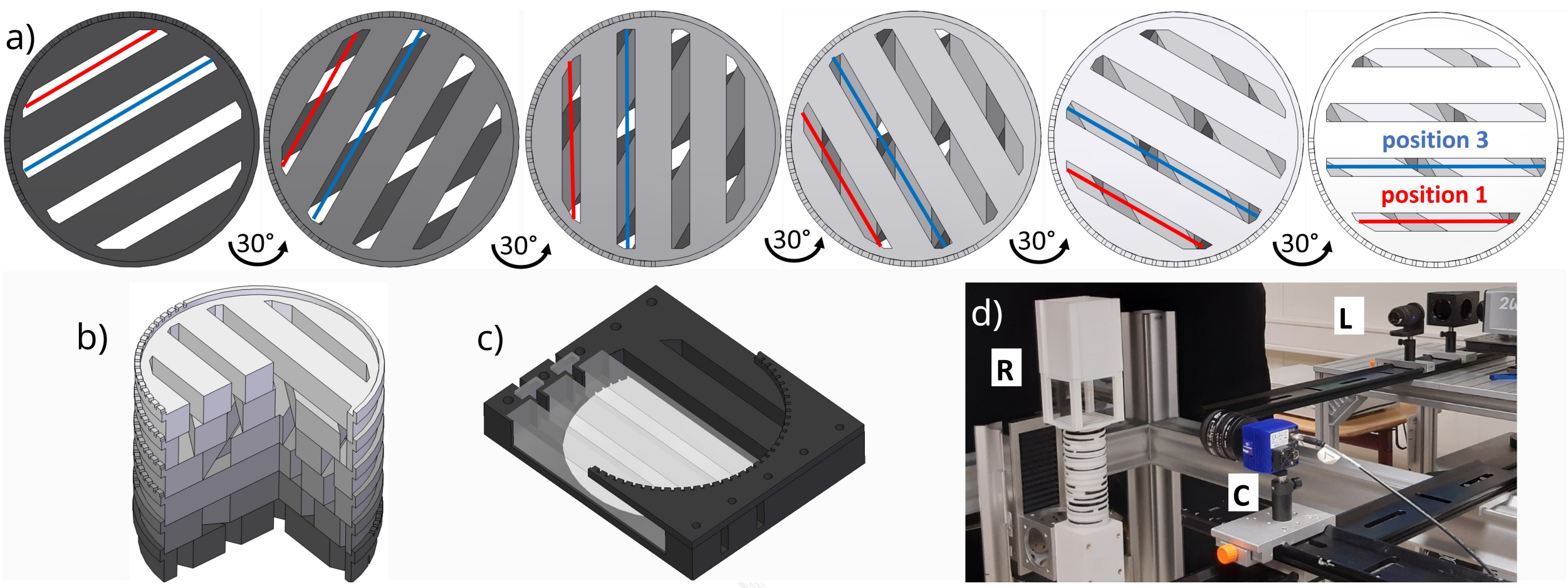

**Figure 1.** a) Schematic of the assembly process of the packed bed using $30°$ rotation between modules. The red and blue lines mark the two measurement positions (P1 and P3) in different layers. b) The cutout illustrating the structure of void spaces in the bed. c) An optical module. d) The PIV setup including the packed bed and the measuring equipment.

### 2.1 Experimental setup

To allow optical access inside the packed bed, a standard module is replaced by an optical one (Fig. 1 c), where opaque bars are exchanged by transparent fused silica rods. The full assembly of the packed-bed reactor (R) is presented in Fig. 1 (d). Airflow is controlled by a Bronkhorst mass flow controller, enters the bed through a diffuser to homogenize the flow, and is seeded by Di-Ethyl-Hexyl-Sebacat (DEHS) tracer. The measurement region is illuminated by a double pulsed Quantel Q-smart Twins 850 Nd:YAG PIV-laser (L) and an Imager CX 12MP Camera (C) from LaVision GmbH equipped with a Nikon AF Nikkor 35 mm f/2D lens records the PIV signal. The flow data was averaged over 500 images, recorded with 10 Hz (50 s physical time). A detailed description of the experimental apparatus and related datasets can be found in [22].

The coordinate system of the packed bed is defined so that the $z$ axis is going through the center of rotation of each module, with the origin of the coordinate system placed in the plane of the bottom surface of the 1st module. The $x$ axis is parallel to the bars in the topmost module and $y$ coordinate is defined so that the system is right-handed.

The measured data sets are separated into layers, each corresponding to the individual module. The data has been gathered for the layers 13th to 17th and the freeboard region. For each module and the freeboard, the data is gathered in two measurement planes at the mid-span of each void space between the bars, named position 1 (P1) and 3 (P3), and annotated with red and blue lines in Fig. 1 (a).

In the current manuscript only the data from the 17th layer and the freeboard are discussed (for overview of the flow field in the other layers, the reader is referred to [22]). The measurement and simulation data are described in the $(\xi, z)$ coordinate system defined so that the vertical axis is the same as in the global coordinate system, and $\xi$ axis represents the in-plane horizontal coordinate of P1 and P3, with the middle of the chosen plane corresponding to $\xi = 0$.

## 3 Numerical model

### 3.1 Simulations on boundary-conforming mesh

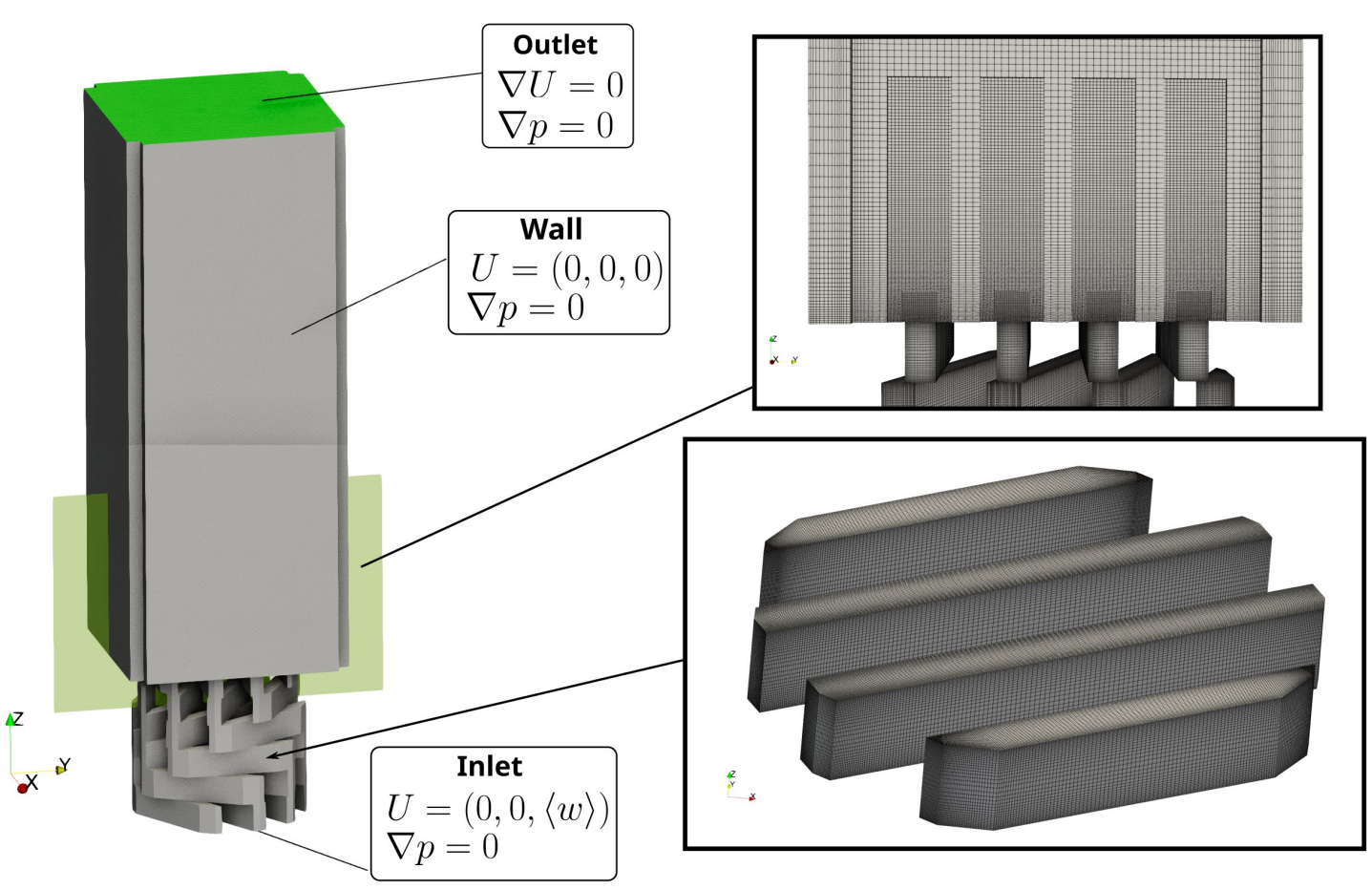

**Figure 2.** Schematic of the computational domain for the simulations on the boundary-conforming mesh along with the boundary conditions prescribed on each wall (left), and the computational grid used for these simulations on the right.

The computations have been performed using *OpenFOAM-12*, using an unsteady solver implementing incompressible Navier–Stokes equations. Based on the indications from our previous work, the flow conditions in the bed are influenced by the freeboard only near the bed surface [6]. Hence, to limit the computational effort, only the top six layers were realized in the simulation (see Fig. 2). The volume of fluid in the modules has been meshed with a boundary conforming hexahedral-dominant structured mesh. The initial mesh for the freeboard region has been generated in the same way, after which it was refined using *snappyHexMesh*. The whole mesh, used for each simulation, consists of around 8.5 million cells. The modules and the freeboard have been coupled using the Non-Conformal Coupling (NCC) interface. The uniform velocity profile with value $\langle w \rangle$ is prescribed at the inlet to the lowest layer, while a reference pressure is set at the outlet. The rest of the boundaries are treated as no-slip walls.

The time step of the simulation is adjusted to keep the Courant number under $0.6$. Both spatial and temporal derivatives are computed using 2nd order schemes. PISO algorithm in the *consistent* variant is used to couple the velocity and pressure, with 5 pressure corrector steps and 1 non-orthogonality correction step. To ensure that the flow is not influenced by the initial conditions, the time-averaging starts after an initial start-up time $T_i$ as listed in Tab 1.

### 3.2 Simulations using blocked-off method

The blocked-off method incorporates the influence of solid boundaries on the flow field by modifying the discretized equations. Patankar [9] originally implemented this via source terms, whereas our approach directly adjusts the system matrix coefficients to avoid complications in residual calculations during the solution of the linear equation system. A general transport equation for a variable $\Phi$, for any cell $c$ with $N$ neighboring cells, can be represented in algebraic form as

$$a_c \Phi_c + \sum_{i=1}^{N} a_i \Phi_i = s_c, \quad (1)$$

with $a_c$, $a_i$ and $s_c$ denoting diagonal, non-diagonal matrix coefficients and source terms, respectively. Cells which are found to be in contact with a solid element are decoupled from the solution process by setting $a_i = 0$, $a_c = 1$, and inserting the desired value to $s_c$, resulting in specifying the cell value and enforcing the boundary condition at the solid-fluid interface.

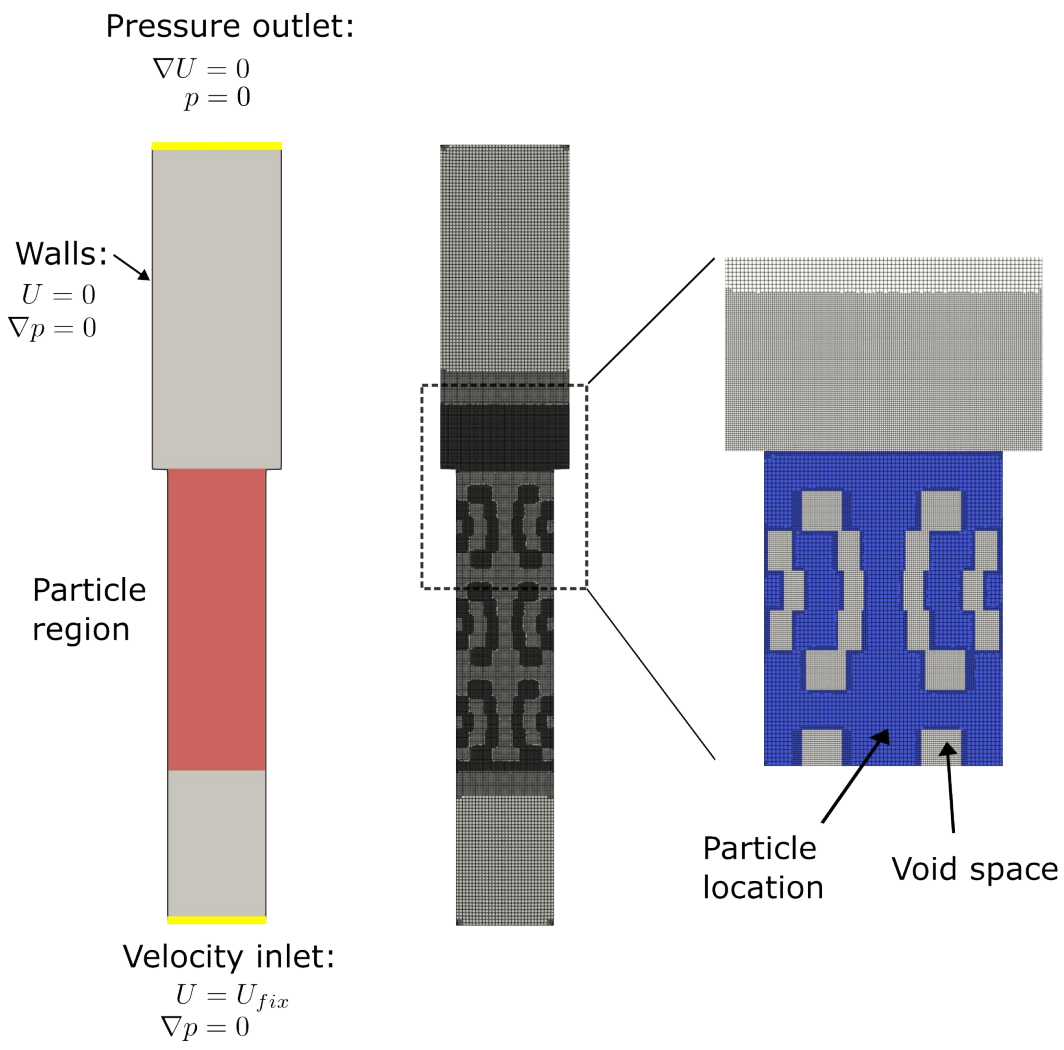

**Figure 3.** Schematic of the computational domain for the simulations employing the blocked-off method along with the boundary conditions prescribed on each wall (left), and the computational grid used for these simulations.

As the particle surface is generally not aligned with the cell faces, the wall shear stress in the control volumes adjacent to the decoupled cells is not represented properly. The correct values of stress are estimated using the fluid's tangential velocity and local distance from the solid surface, and replace the viscous contribution in the momentum equation in these cells. Moreover, the Neumann boundary condition in the pressure equation at the solid surface is enforced by setting relevant $a_i$ coefficients in the pressure equation to a very low value. The current implementation is valid for particles of any shape and is described and validated in [10].

**Table 1.** Times used for data collection in the experiments and both simulation: $T_i$ start-up time, $T_{avg}$ averaging time. The reference time scale is the flow through one module $T_B = B/\langle w \rangle$.

| $Re_p$ | experiment | body-conforming mesh | | blocked-off | |
|---|---|---|---|---|---|
| | $T_{avg}/T_B$ | $T_i/T_B$ | $T_{avg}/T_B$ | $T_i/T_B$ | $T_{avg}/T_B$ |
| 100 | 755 | 12 | 88 | 75.5 | 226.5 |
| 200 | 1515 | 42 | 158 | 151.5 | 454.5 |

The simulations using the blocked-off method are realized with the Bulk-Reaction DEM/CFD framework based on *fireFoam* (*OpenFOAM-v2012*) on a Cartesian grid fitted to the experimental rig using *snappyHexMesh* (see Fig. 3). The dodecagon shape of the walls at the particle bed section is extended downwards defining the inlet. The bars in the test rig are modeled by the introduction of 90 cuboid particles imposed onto the fluid domain via the blocked-off method. Mesh refinement is applied in the void spaces as well as up- and downstream of the particle region, up to the cell size $B/\Delta x = 20$, leading to a total number of 5.1 million cells. The same mesh is used for all Reynolds numbers. The Courant number is kept between values of $0.1$ - $0.3$. The PISO algorithm is utilized to solve the evolution of the flow field with one non-orthogonal corrector step to account for the polyhedral cells at the walls of the test rig. In general, 2nd-order schemes are applied in the discretization of the transport equations, while for convection a 2nd-order upwind scheme is employed to prevent pressure oscillations. Sampling times for averaging are reported in Tab. 1.

# 4 Results

## 4.1 Flow inside the bed

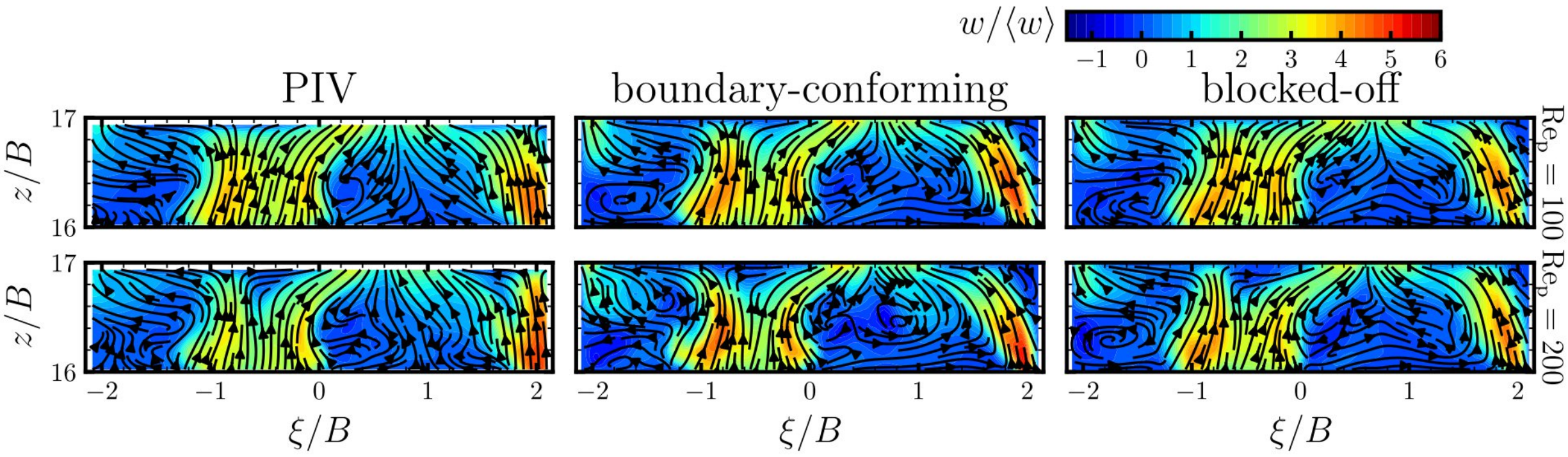

**Figure 4.** Averaged velocity fields at P1 inside layer 17, visualized using streamlines. Color denotes the vertical component $w/\langle w\rangle$. Plotted data has been gathered using PIV and simulation approaches: boundary-conforming and blocked-off methods, at $\mathrm{Re}_p = 100$ (top) and 200 (bottom).

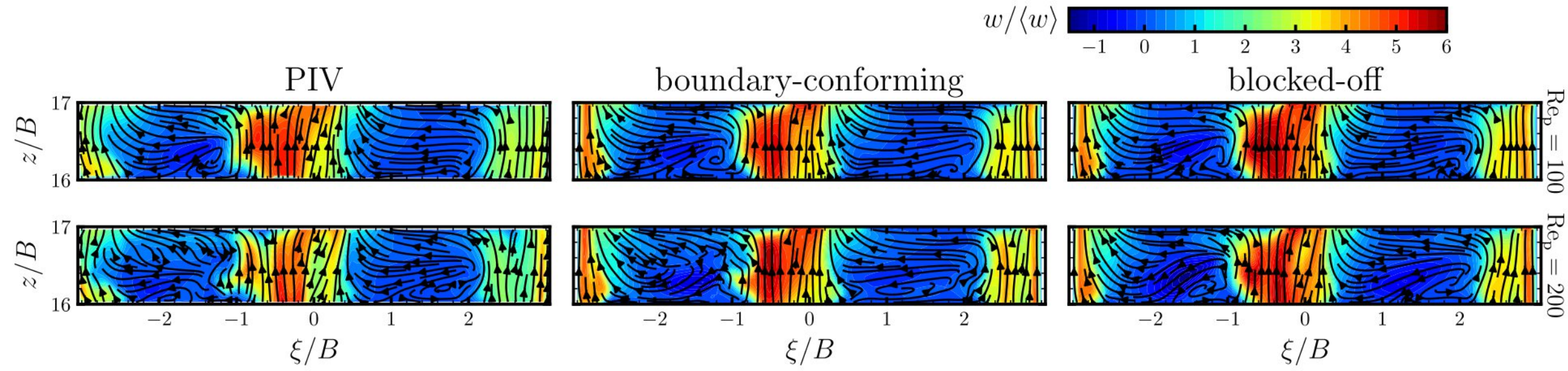

**Figure 5.** Averaged velocity fields at P3 inside layer 17, visualized using streamlines. Color denotes the vertical component $w/\langle w\rangle$. Plotted data has been gathered using PIV and simulation approaches: boundary-conforming and blocked-off methods, at $\mathrm{Re}_p = 100$ (top) and 200 (bottom).

The flow inside the packed bed is visualized in detail in the Figs. 2 and 3, illustrating the velocity field at positions P1 and P3, respectively. At P1, the void space has two inlets from the layers below and two outlets into the above module, which are staggered with respect to the inlets due to the modules' rotation. On the other hand, the P3 interstice has three inlets and outlets. The velocity field is mainly determined by the bed geometry and the connectivity between the void spaces in the layers. The fluid is accelerated in the inlets and its downstream behavior is determined by the positions of the outlets. In case of P1, the flow impinges on the top wall of the module and is forced to split into two directions towards the outlets. Since the left outlet has much smaller cross-sectional area, most of the flow converges from both inlets into the right outlet. Position P3 is closer to the axis of rotation, hence, the outlets are more aligned with the inlets and a strong upwards flow between them can be observed. The regions positioned between the inlet/outlet pairs are characterized by a lateral flow, presumably driven by the pressure difference between these regions. This behavior also leads to the presence of recirculation regions, attached to the left side of the in-flowing streams.

The simulation results agree very well with the PIV measurements. The main differences can be observed in the far left and right areas of the measured regions, where, the simulation data indicates the presence of the boundary layer jets for P3 and recirculation regions attached to the bottom left and top right corners of P1. On the other hand, based on streamline visualization, the PIV data indicates the fluid crossing the boundary of the measured region on the left side of P1 and P3 for both $\mathrm{Re}_p$. This effect can be attributed to a different shape of the void spaces in the optical module (Fig. 1 c). To allow for optical access, the

void spaces are longer to allow for a perpendicular laser inlet face, and an additional space on the left and right of the measured region is characterized by a cavity like flow. The impact of the geometry modification on the flow field is small and remains localized to the near wall regions.

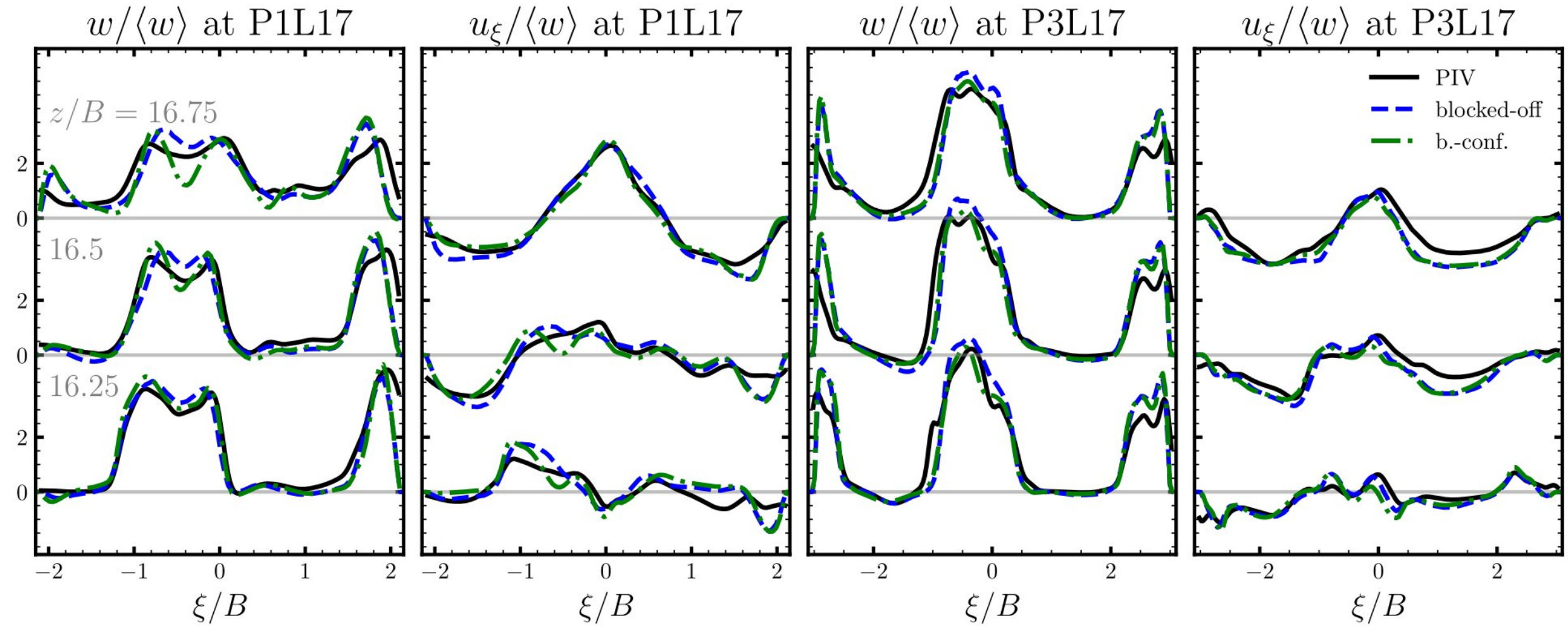

**Figure 6.** Averaged vertical ($w$) and horizontal ($u_\xi$) velocity components inside the bed for $\mathrm{Re}_p = 100$, plotted along the lines at vertical coordinates $z/B = 16.25, 16.5, 16.75$ (annotated by gray numbers) in planes P1 (left) and P3 (right) in the 17th layer.

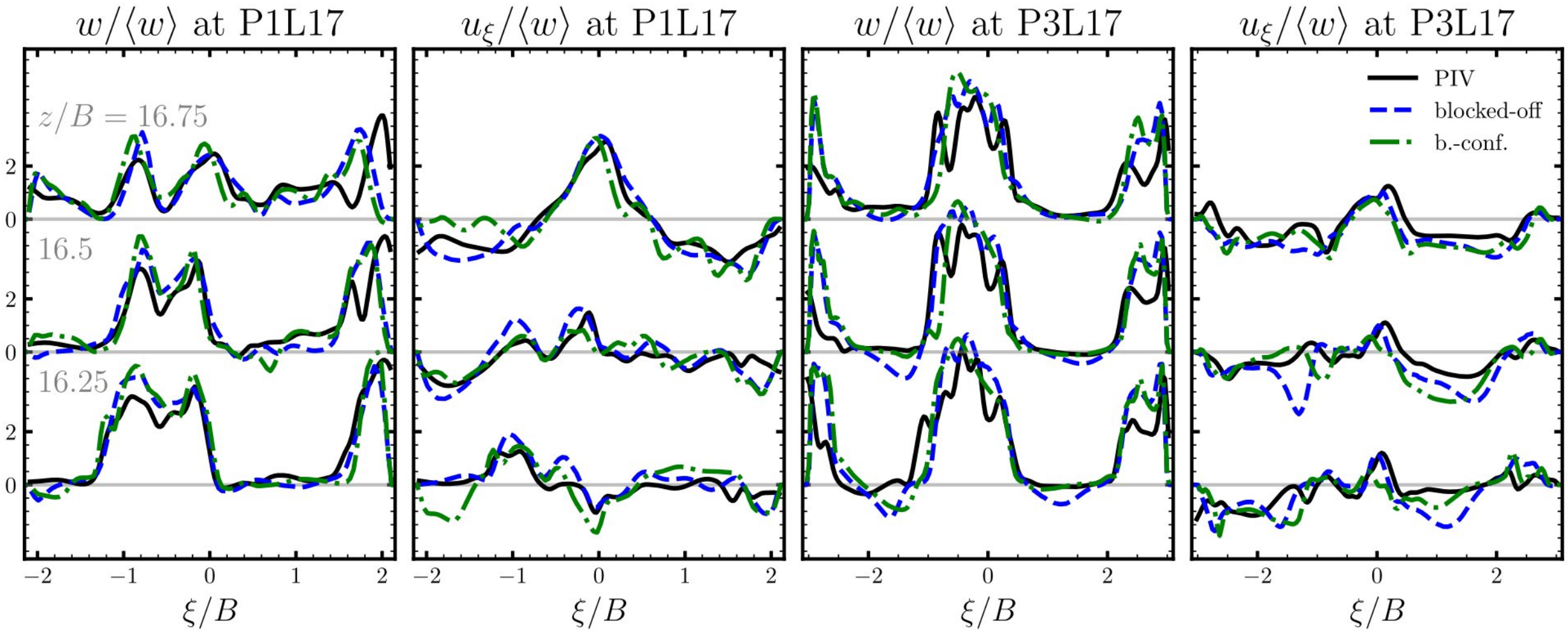

**Figure 7.** Averaged vertical ($w$) and horizontal ($u_\xi$) velocity components inside the bed for $\mathrm{Re}_p = 200$, plotted along the lines at vertical coordinates $z/B = 16.25, 16.5, 16.75$ (annotated by gray numbers) in planes P1 (left) and P3 (right) in the 17th layer.

The differences in the flow field near the side walls, are further presented quantitatively in Figs. 6 and 7 corresponding to $\mathrm{Re}_p = 100$ and $200$, respectively. Both simulation approaches predict a different shape of the boundary layer jets (at P1 on the right and at P3 for both sides of the measurement region), with the simulations consistently overpredicting the maximum streamwise velocity in these regions. At the larger Reynolds number (Fig. 7), the differences between the measured and simulated flow fields increase.

Although, the flow fields are largely similar across the two Reynolds numbers, the increase in the fluid's inertia influences the shapes and sizes of the separated flow regions. At P1, the increase of Reynolds number leads to a much more complex flow field between the two inlets, with recirculating flow appearing in the middle of the measurement section. At the other position, larger $\mathrm{Re}_p$ leads to increased separation on the right side of the measured region. On the left side, however, it seems that a faster reat-tachment can be observed. PIV data plotted in Fig. 6 and 7 also indicates that the non-uniformity of the flow field is much stronger at larger $\mathrm{Re}_p$, although this effect is not fully represented in the simulation data.

In general, the numerical methods align very well in the void space with only minor differences in height and shape of the jets. Though the boundary-conforming and blocked-off approach yield different mesh refinements close to walls/particles (typical near-wall boundary layer refinement vs. equally spaced mesh), the line plots (Figs. 4 and 5) show almost no difference in the prediction of the wall jets for $\mathrm{Re}_p = 100$ and only slight differences for larger $\mathrm{Re}_p$, which are not significant, in view of the deviations between simulations and experiment. The strongest difference between the two simulation approaches are found for $u_\xi$ at P1 for $\mathrm{Re}_p = 200$ where a more pronounced horizontal motion in the recirculation region is predicted by the boundary-conforming method in contrast to the blocked-off approach and the experiment.

The additional assessment of the overall performance of the blocked-off method *vs.* the boundary-resolving DNS, can be achieved by comparison of the pressure drop $\Delta p$ over the length of the bed $L$, which can be expressed in a non-dimensional form as a friction factor $f = -B\,\Delta p/(\langle w\rangle_s^2 L)$. As the present study only considers a packed bed with a reduced number of layers using the boundary-conforming approach (see section 3.1), we leverage the findings of Demir et al. [20] who evaluated the friction factor for the considered packed in in a fully periodic configuration in a streamwise direction (ruling out the influence of inlet or freeboard conditions) using boundary-conforming method. They reported values of $f = 75.03$ and $72.87$ for Reynolds numbers $100$ and $200$, respectively. The blocked-off method results in the values of $f_{bo} = 77.68$ and $71.52$ for the same Reynolds numbers, yielding relative absolute differences of around $3.5\%$ and $1\%$ which is within the margin of error found between coarse and fine mesh resolution of the same configuration in [20]. A reader interested in the distribution of the pressure fields from the blocked-off simulations is referred to the supplementary material of this article (Fig. S1). Based on these macroscopic results and microscopic prediction of the flow field, the handling of the wall flow with the blocked-off approach seems to be appropriate for the range of studied Reynolds numbers.

### 4.2 Freeboard

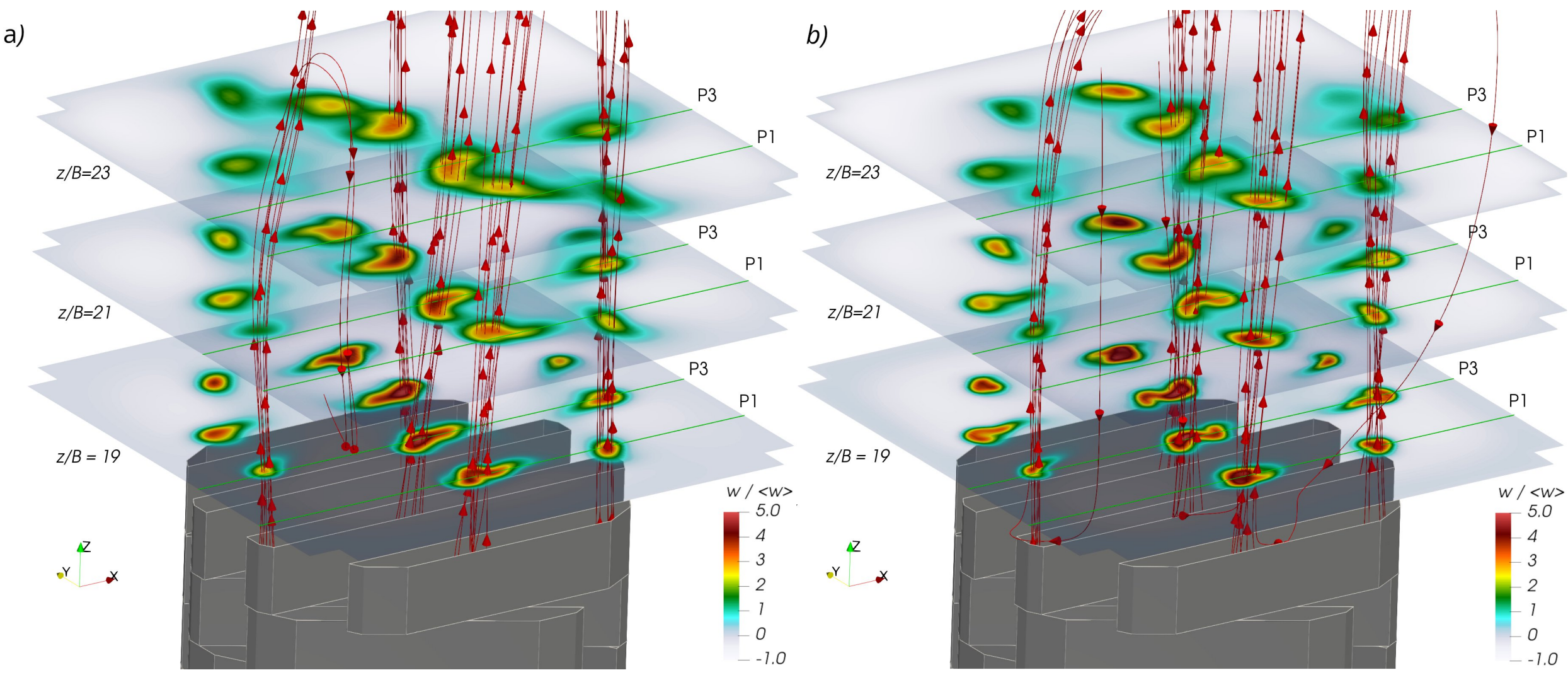

**Figure 8.** An overview of the flow in the freeboard: velocity fields from the simulations employing boundary-conforming mesh for $\mathrm{Re}_p = 100$ (a) and 200 (b) at the planes at $z/B = 19$, 21 and 23. Colormap denotes the vertical velocity component $w/\langle w\rangle$ and the red lines visualize the streamlines of the flow field in a selection of jets. Green lines indicate sampling lines at the positions P1 and P3.

The freeboard flow field, shown in Fig. 8, is dominated by jets emerging from the outlets of the top module—two jets in plane P1 and three in P3. The velocity field is complex; as the distance from the bed surface increases, the jets dissipate and interact with each other, which results in merging of the jets issued from positions P1 and P3. This is especially visible for the lower Reynolds number (Fig. 8a), as the increased inertial effects for $Re_p = 200$ lead to less diffuse jet shapes. Moreover, the streamlines visualization indicates that part of the fluid ejected from the bed is redirected towards the bed surface, hinting at the presence of large recirculation regions surrounding the jets.

The main difference between both Reynolds numbers relates to the nature of the flow above the bed. For the lower Reynolds number, the jets and the flow downstream remain stationary. On the other hand, the increase of $\mathrm{Re}_p$ leads to unsteady flow above the bed, with the jets slowly oscillating in lateral directions, which leads to three-dimensional vortex shedding downstream.

These flow features can be further observed in detail in Fig. 9 comparing all three datasets at planes located at P1 and P3. Although, the jets themselves are more pronounced near the bed surface for higher $\mathrm{Re}_p$ due to stronger inertial effects (e.g., the shape of the leftmost jet at P3 is preserved longer), the unsteady flow results in a more diffusive behavior, and a quicker dissipation of the jets downstream (best exemplified by the jets in plane P1). At $\mathrm{Re}_p = 100$, the jets issued at P1 seem to merge rapidly above the bed surface. A comparison with Fig. 8 indicates, that the left jet shifts out of plane P1, merging with the P3 jet, which results in an apparent slowdown of the fluids velocity in Fig. 9.

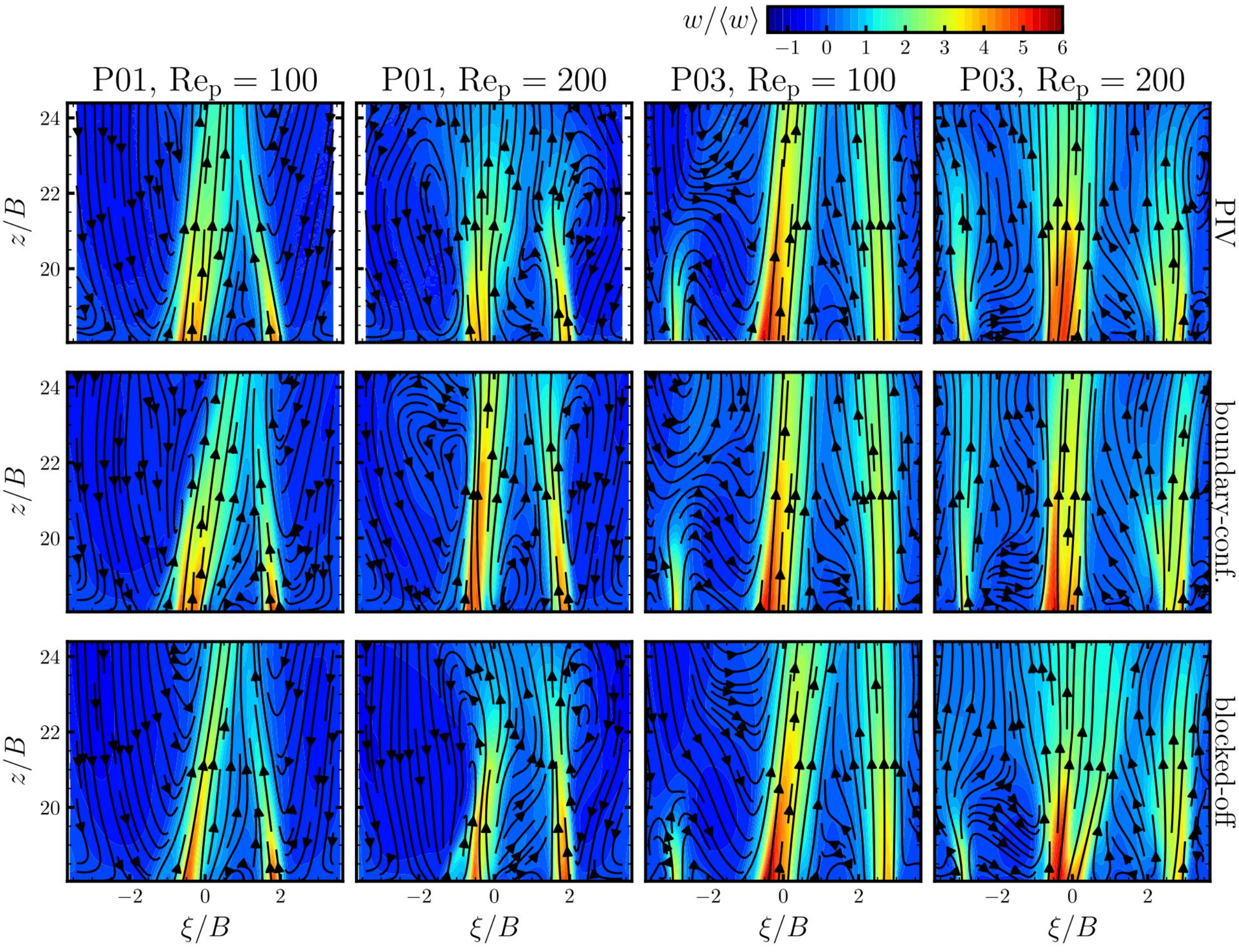

**Figure 9.** Velocity field in the freeboard for $\mathrm{Re}_p = 100$ and 200 at P1 (left) and P3 (right), visualized using streamlines. Colormap denotes the vertical component $w/\langle w\rangle$. Plotted data has been gathered using PIV and simulation approaches: boundary-conforming and blocked-off methods.

At the sides of the P1 plane at $\mathrm{Re}_p = 100$, slow-moving gas is drawn from the outlet into the freeboard and below the bed surface, located at $z/B = 18$, forming two large recirculation zones along the channel walls which are driven by the jets. In the P3 plane, the backflow is weaker and only present due to the leftmost jet impinging onto the descending fluid which, in turn, redirects the jet towards the bed surface and into the middle jet. The effect of this *backflow* into the reactor could be crucial when considering scalar mixing in such systems as it might reintroduce the exiting gas back into the packed bed. At the higher $\mathrm{Re}_p$, the unsteady flow helps redistribute the jets' momentum laterally, yielding a more uniform velocity field and smaller recirculation regions in the middle of the freeboard.

Both, simulations and experimental measurements are in good agreement regarding the prediction of the flow over the bed surface, especially for the lower Reynolds number case. Each data set predicts very similar streamwise and horizontal velocities, plotted in Fig. 10 and 11.

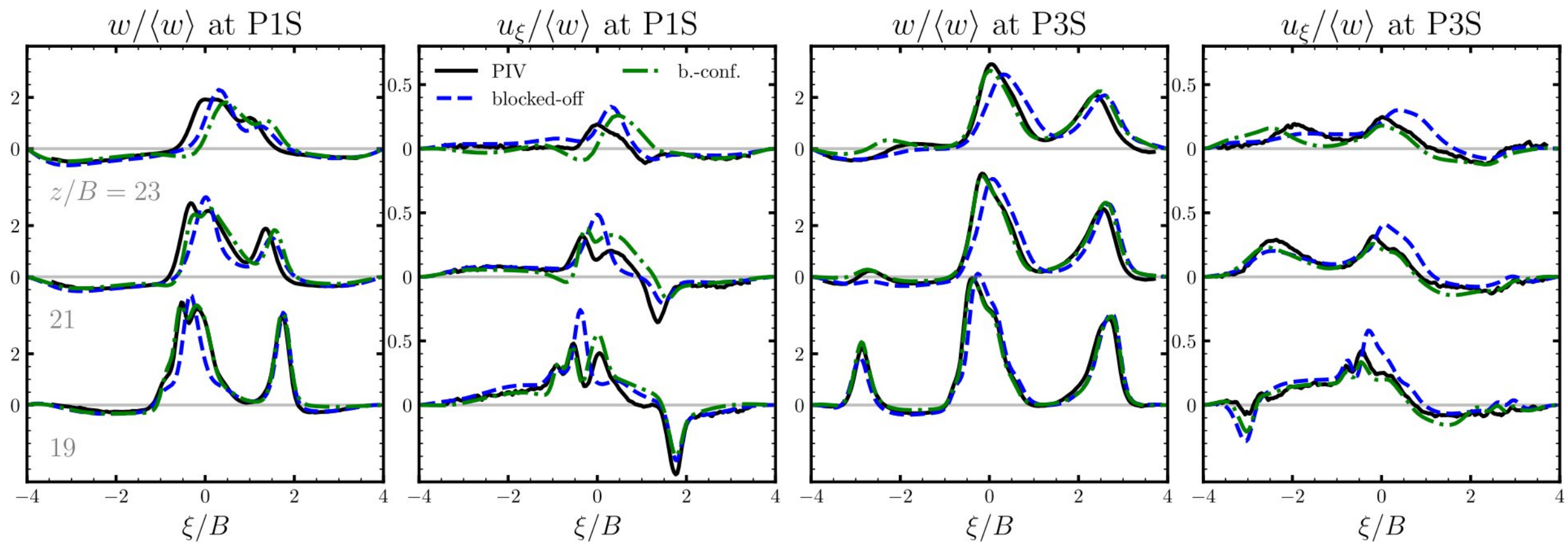

**Figure 10.** Averaged vertical ($w$) and horizontal ($u_\xi$) velocity components in the freeboard for $\mathrm{Re}_p = 100$, plotted along the lines at vertical coordinates $z/B = 19, 21, 23$ (annotated by gray numbers) in planes P1 (left) and P3 (right).

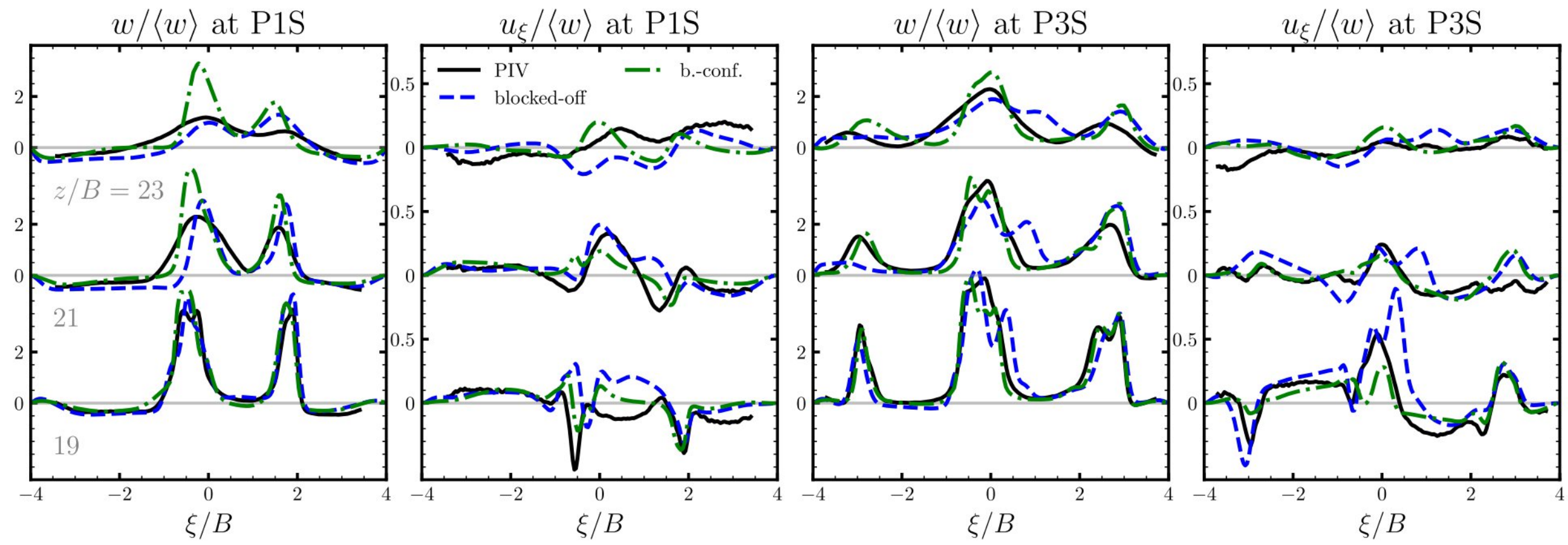

**Figure 11.** Averaged vertical ($w$) and horizontal ($u_\xi$) velocity components in the freeboard for $\mathrm{Re}_p = 200$, plotted along the lines at vertical coordinates $z/B = 19, 21, 23$ (annotated by gray numbers) in planes P1 (left) and P3 (right).

Examining first the results for lower $\mathrm{Re}_p$, near the bed surface (at $z/B = 19$), the boundary-conforming method reproduces the measured vertical velocity $w$ and shapes of the jets very accurately, while the blocked-off method predicts a too narrow jet on the left of P1. Similarly, the data from the blocked-off simulation overpredicts $w$ in the center jet at P3, which also leads to a bigger discrepancy in the horizontal velocity component. We attribute these small differences to the accuracy of representation of the bed geometry by the IBM: the stress distribution in the void spaces issuing the jets determines to a large extent the direction and shape of the jet downstream. Additionally, the mesh resolution at the walls is in general lower compared to the simulations employing body-conforming mesh which reduces the accuracy of the representation of the near-wall gradients. This might be increasingly important near the bed surface where the flow complexity increases. Therefore, further mesh refinement in these regions would presumably yield improved results.

With increasing bed height, differences between simulations and the experiment get more pronounced. Here, the three-dimensional interaction of the jets comes into play, causing small discrepancies in the jets to accumulate to larger deviations from experimental data. The errors in momentum prediction are probably caused by a limited resolution downstream of the bed surface diffusing and dampening the breakdown of the jets into smaller vortical structures. Moreover, insufficient resolution and therefore inaccurately captured interactions between the jets could lead to the a shift in their position perpendicular to the planes shown in Fig. 9, which would be strongly reflected in the plotted profiles. Alternatively, different boundary conditions at the domain outlet (fixed pressure condition for simulations vs. small environmental interference in the experiment) may also contribute to different velocity fields above the bed.

Following the line plots of the higher Reynolds number in Fig. 11, the agreement of the numerical data with the PIV data is again best close to the bed surface for P1. In the experiment and the blocked-off simulation the mean jets dissipate faster than in the body-conforming case. For P3, the outer jets are reproduced well but the center jet shows a different flow structure with two velocity peaks in the numerical data for $z/B = 19$ and $21$. Importantly, the accuracy of the simulation data might be limited by a relatively short averaging time (especially in case of the boundary-conforming method). Previous work [6] indicated that in such systems, exceedingly long averaging times are required to fully capture the dynamics of the freeboard. Although the convergence of statistics has to be further assessed, the overall nature of the flow field is reproduced correctly by both simulation approaches.

## 5 Conclusions & Outlook

This work examines the flow through a modular packed bed of bar-like particles using PIV measurements and simulations with both boundary-conforming and blocked-off approaches. Two Reynolds numbers, $\mathrm{Re}_p = 100$ and $200$, are considered, and the flow behavior is analyzed both inside and above the bed surface.

The study improves the current understanding of flow dynamics within and above the packed bed of non-spherical particles. Inside the bed, the flow features are largely determined by the geometry and alignment of inlets and outlets in each void space, with the Reynolds number increase having only a limited effect on the flow field structure. The complex geometry of the void spaces results in a strongly non-uniform velocity field and the presence of multiple separated regions. Above the bed surface, the fluid behavior is primarily influenced by the strong jets emanating from the surface-adjacent void spaces, driving large recirculation regions. At $\mathrm{Re}_p = 200$, the flow becomes unsteady, leading to faster dissipation of the jets and a more uniform flow field above the reactor.

Additionally, based on the results of the simulations conducted on the boundary-conforming mesh, less than six particle layers (the bed consists of 18 layers) are required for the flow to become independent of the inlet conditions and correspond to the measured velocity field. This finding indicates that future studies of flow phenomena in this and similar geometries may represent only a section of a packed bed in the simulation without accuracy loss, enabling reduction of computational cost.

While the flow inside the packed bed shows good agreement between experiments and simulations, differences arise due to necessary geometry adjustments for measuring the flow field between particles. More pronounced deviations are observed in the freeboard region, where boundary-conforming simulations align well with experiments at the bed surface, but discrepancies in jet dissipation rates into the freeboard persist. The blocked-off approach yields better agreement in this region, though mismatches between emerging jets remain, potentially due to insufficient mesh refinement in the freeboard.

A key contribution of this work is the validation of the blocked-off approach for such geometries. The method, while computationally inexpensive and easy to implement, demonstrates good agreement with experimental data and effectively captures all flow features inside the bed, confirming its capability to model flow through packed beds of polyhedral particles. Thus, it provides a practical alternative to more sophisticated Immersed Boundary Methods (IBM) for capturing the main flow features inside packed beds. On the other hand, the use of Non-Conformal Coupling interfaces with the boundary-conforming mesh has produced stable and accurate solutions, paving the way for future studies of non-static packed beds with moving particles.

Comparison of simulation results with experimental measurements highlights the difficulties in predicting accurately the flow of fluid near the surface of the packed bed. This region is more challenging to accurately simulate than the interstitial space, and the complexity increases with the Reynolds number. Based on obtained results the flow field prediction near the surface is highly sensitive to the choice of the boundary treatment method and the mesh resolution.

Further investigations are needed to fully assess the accuracy of both simulation approaches, focusing on fluctuations in the freeboard, turbulent kinetic energy, and the influence of averaging time on the results. Exploring higher Reynolds number cases would also provide valuable insights into industrial packed bed conditions.

**Supporting Information**

Supporting Information for this article can be found under [Link provided by Wiley]. This section includes the distribution of pressure fields inside the bed computed using the blocked-off method.

**Acknowledgment**

This work was funded by the Deutsche Forschungsgemeinschaft (DFG, German Research Foundation) – Project-ID 422037413 - TRR 287 and the experimental equipment by DFG project number 279416000.

**Symbols used**

$A$ [$m^2$] bed cross-section area

$B$ [m] bar size

$\mathrm{Re}_p$ [-] particle Reynolds number

$T$ [s] time

$\langle w \rangle$ [$ms^{-1}$] bulk velocity

$w$ [$ms^{-1}$] vertical velocity

$u_\xi$ [$ms^{-1}$] horizontal velocity along the PIV measurement plane

$Q$ [$m^3s^{-1}$] volumetric flow rate

$x,y,z$ [m] PIV coordinate system coordinates

*Greek letters*

$\alpha$ [-] rotation angle

$\phi$ [-] porosity

$\xi$ [m] horizontal coordinate along the PIV measurement plane

$\nu$ [$m^2s^{-1}$] viscosity

*Sub- and Superscripts*

| | |
|---|---|
| $i$ | *initialization* |
| $avg$ | *averaging* |
| $c$ | *cell variable* |

*Abbreviations*

| | |
|---|---|
| *P1* | *measurement position 1 (see Fig. 1)* |
| *P3* | *measurement position 3 (see Fig. 1)* |
| *DEHS* | Di-Ethyl-Hexyl-Sebacat |
| *NCC* | Non-Conformal Coupling |
| *PIV* | Particle Image Velocimetry |
| *CFD* | Computational Fluid Dynamics |
| *IBM* | Immersed Boundary Method |
| *FVM* | Finite Volume Method |

**Table and Figure captions**

**Table 1.** Times used for data collection in the experiments and both simulation: $T_i$ start-up time, $T_{avg}$ averaging time. The reference time scale is the flow through one module $T_B = B/\langle w \rangle$.

**Figure 1.** a) Schematic of the assembly process of the packed bed using $30°$ rotation between modules. The red and blue lines mark the two measurement positions (P1 and P3) in different layers. b) The cut-out illustrating the structure of void spaces in the bed. c) An optical module. d) The PIV setup including the packed bed and the measuring equipment.

**Figure 2.** Schematic of the computational domain for the simulations on the boundary conforming mesh along with the boundary conditions prescribed on each wall (left), and the computational grid used for these simulations on the right.

**Figure 3.** Schematic of the computational domain for the simulations employing the blocked-off method along with the boundary conditions prescribed on each wall (left), and the computational grid used for these simulations.

**Figure 4.** Averaged velocity fields at P1 inside layer 17, visualized using streamlines. Color denotes the vertical component $w/\langle w \rangle$. Plotted data has been gathered using PIV and simulation approaches: boundary-conforming and blocked-off methods, at $\mathrm{Re}_p = 100$ (top) and 200 (bottom).

**Figure 5.** Averaged velocity fields at P3 inside layer 17, visualized using streamlines. Color denotes the vertical component $w/\langle w \rangle$. Plotted data has been gathered using PIV and simulation approaches: boundary-conforming and blocked-off methods, at $\mathrm{Re}_p = 100$ (top) and 200 (bottom).

**Figure 6.** Averaged vertical ($w$) and horizontal ($u_\xi$) velocity components inside the bed for $\mathrm{Re}_p = 100$, plotted along the lines at vertical coordinates $z/B = 16.25, 16.5, 16.75$ (annotated by gray numbers) in planes P1 (left) and P3 (right) in the 17th layer.

**Figure 7.** Averaged vertical ($w$) and horizontal ($u_\xi$) velocity components inside the bed for $\mathrm{Re}_p = 200$, plotted along the lines at vertical coordinates $z/B = 16.25, 16.5, 16.75$ (annotated by gray numbers) in planes P1 (left) and P3 (right) in the 17th layer.

**Figure 8.** An overview of the flow in the freeboard: velocity fields from the simulations employing boundary-conforming mesh for $\mathrm{Re}_p = 100$ (a) and $200$ (b) at the planes at $z/B = 19$, $21$ and $23$. Colormap denotes the vertical velocity component $w/\langle w \rangle$ and the red lines visualize the streamlines of the flow field in a selection of jets. Green lines indicate sampling lines at the positions P1 and P3.

**Figure 9.** Velocity field in the freeboard for $\mathrm{Re}_p = 100$ and $200$ at P1 (left) and P3 (right), visualized using streamlines. Colormap denotes the vertical component $w/\langle w \rangle$. Plotted data has been gathered using PIV and simulation approaches: boundary-conforming and blocked-off methods.

**Figure 10.** Averaged vertical $(w)$ and horizontal $(u_\xi)$ velocity components in the freeboard for $\mathrm{Re}_p = 100$, plotted along the lines at vertical coordinates $z/B = 19, 21, 23$ (annotated by gray numbers) in planes P1 (left) and P3 (right).

**Figure 11.** Averaged vertical $(w)$ and horizontal $(u_\xi)$ velocity components in the freeboard for $\mathrm{Re}_p = 200$, plotted along the lines at vertical coordinates $z/B = 19, 21, 23$ (annotated by gray numbers) in planes P1 (left) and P3 (right).

## Entry for the Table of Contents

**Research Article:** This study examines gas flow in a modular packed bed reactor with square bars arranged in rotated layers, creating complex void spaces. Using Particle Image Velocimetry, measurements were taken at Reynolds numbers of 100 and 200. These results validated two particle-resolved numerical simulations employing boundary-conforming meshing and blocked-off methods. Flow within the bed is geometry-dependent, while the freeboard exhibits dissipating jets with unsteady oscillations at higher Reynolds numbers. Both numerical methods align well with measurements, though deviations in the freeboard are attributed to simulation limitations.

**Low Reynolds number flow in a packed bed of rotated bars**

W. Sadowski, C. Velten, M. Brömmer, H. Demir, K. Hülz, F. di Mare, K. Zähringer, V. Scherer

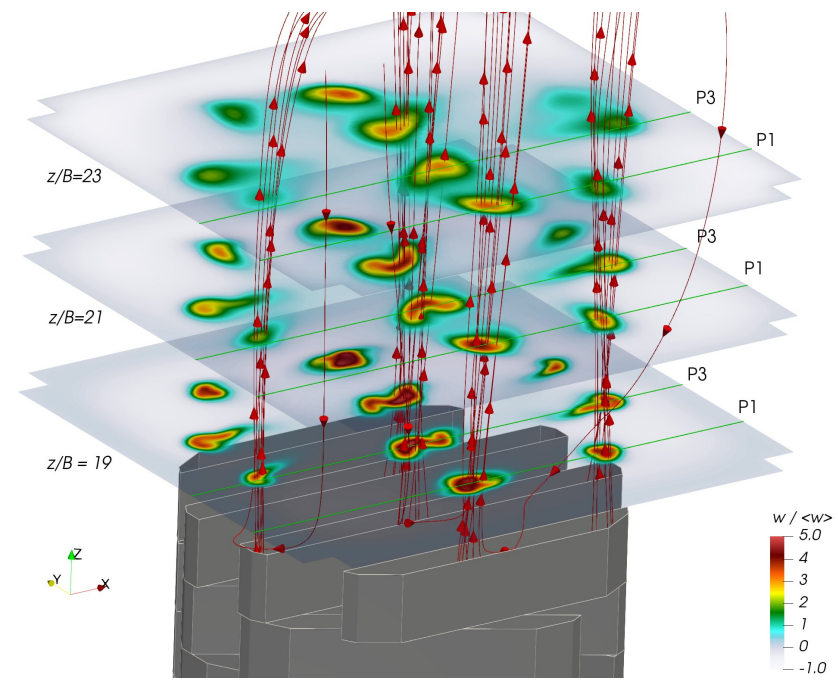